\documentclass{llncs}

\usepackage[pdftex]{graphicx}

\usepackage[font=footnotesize]{subfig}
\usepackage{color}

\begin{document}

\newcommand{\revised}[1]{{\color{black}#1}}

\title{A Holistic Approach to Log Data Analysis in High-Performance Computing Systems:\\The Case of IBM Blue Gene/Q}

\titlerunning{Blue Gene/Q data analysis}

\author{Alina S\^irbu \and  Ozalp Babaoglu}
\authorrunning{A. S\^irbu \and  O. Babaoglu}
\tocauthor{Alina S\^irbu and  Ozalp Babaoglu}

\institute{Department of Computer Science and Engineering, University of Bologna,\\
Mura Anteo Zamboni 7, 40126 Bologna, Italy\\
\email {alina.sirbu@unibo.it,ozalp.babaoglu@unibo.it}}

% make the title area
\maketitle

\begin{abstract}
The complexity and cost of managing high-performance computing infrastructures are on the rise. Automating management and repair through predictive models to minimize human interventions is an attempt to increase system availability and contain these costs. Building predictive models that are accurate enough to be useful in automatic management cannot be based on restricted log data from subsystems but requires a holistic approach to data analysis from disparate sources. Here we provide a detailed multi-scale characterization study based on four datasets reporting power consumption, temperature, workload, and hardware/software events for an IBM Blue Gene/Q installation. We show that the system runs a rich parallel workload, with low correlation among its components in terms of temperature and power, but higher correlation in terms of events. As expected, power and temperature correlate strongly, while events display negative correlations with load and power. Power and workload show moderate correlations, and only at the scale of components. The aim of the study is a systematic, integrated characterization of the computing infrastructure and discovery of correlation sources and levels to serve as basis for future predictive modeling efforts. 
\end{abstract}

%%%%%%%%%%%%%%%%%%%%%%%%%%%%%%%%%%%%%%%%%%%%%
%%%%%%%%%%%%%%%%%%%%%%%%%%%%%%%%%%%%%%%%%%%%%
%%%%%%%%%%%%%%%%%%%%%%%%%%%%%%%%%%%%%%%%%%%%%
\section{Introduction} 

%%%%%%%%%%%%%%%%%%%%%%%%%%%%%%%%%%%%%%%%

As the size and complexity of high-performance computing (HPC) infrastructures continue to grow driven by exascale speed goals, maintaining reliability and operability levels high, while keeping management costs low, is becoming increasingly challenging.  Continued reliance on human operators for management and repair is not only unsustainable, it is actually detrimental to system availability:  in very large and complex settings like data centers, accidental human errors have been observed to rank second only to power system failures as the most common causes of system outages~\cite{Pone2013}.

Large computing systems produce large amounts of data in the form of logs tracing resource consumption, errors, events, etc. These data can be put to use for understanding system behavior and for building predictive models to tackle the management challenges. Most studies in this direction have focused on particular subsystems rather than the system as a whole, which is a necessary condition for achieving realistic models with good predictability traits~\cite{Salfner2010}.  With recent progress in \emph{Data Science} and \emph{Big Data}, it is becoming increasingly feasible to carry out such a holistic analysis towards improving predictions by considering data from a variety of sources covering different subsystems and measures~\cite{Gao2014}. 

%%REVIEWER 2:
%%%Motivation:  The seems to position this work as important for the further prediction work, but the authors do not clearly motivate why this work is so important to the end goal and how this work differs from all the other failure prediction and correlation work currently out there.

%%AS:
%%%%I added a statement in the next paragraph saying this analysis is a first step to feature identification.

In this paper we report the results of a characterization study integrating four datasets from different subsystems in an effort to understand the behavior of a 10-rack IBM Blue Gene/Q~\cite{lakner2013} installation and
quantify the correlations among power, temperature, workload, and hardware/software events as well as among different system components. In certain cases, we report the lack of correlations, which can be just as important as their presence. \revised{These results provide a first step towards identifying important features for future predictive studies.}

%%REVIEWER 2:
%%%Novelty: The novelty of the approach taken in this work is not adequately addressed.  The only solid results presented, correlation between temp and power, is known to the community.  Looking at environmental and failure data for trends is has also been done before, as pointed out by the authors.  How does this work add to the field?  The paper does not adequately answer this question.

 %%REVIEWER 3:
%%%Overall, the paper is lacking novelty.  Using correlation is very natural approach to understand the system behavior, and this approach can be found in many papers. Although several ideas are proposed at the end of the paper, the authors do not propose any methodologies/tools (beyond simple correlations) for practical merit.

%%AS:
%%%I added a statement in the next paragraph to underline the integration of many datasets, which is the main difference from other studies, and main novelty. I also added a few statements to the related work section to underline this novelty.  I hope this addresses at least in part the 2 comments.

The contributions of this paper are threefold. First, we provide a characterization of a Blue Gene/Q system from thermal, power, workload, and event log perspectives, highlighting significant features for system behavior and the presence and absence of correlations between different components.
No correlation in terms of power and thermal behavior was found across components, yet events exhibit significant spatial correlations, indicating possible propagation of errors. Secondly, an integrated analysis of the four datasets searches for correlations among various metrics so as to identify further possible relations for future modeling and prediction studies. This reveals significant positive correlation between power consumption and temperature, and a weaker negative correlation between hardware/software events and power or workload. There are also indications of correlations between workload and power but only at a finer spatial granularity (at rack rather than at system scale). Thirdly, we use the preliminary indications on the importance of different features for explaining system behavior to propose a feature set to be used in future work for event prediction. \revised{An important feature of our study is its holistic nature integrating multiple datasets, to an extent not present in the literature, neither in terms of system characterization, nor in terms of correlation and predictive studies.}

In the next section we describe the data, while Section~\ref{data_analysis} contains our analysis for individual and integrated datasets. Section~\ref{related} includes related work, and Section~\ref{discussion} discusses future predictive studies and data quality.
%%%%%%%%%%%%%%%%%%%%%%%%%%%%%%%%%%%%%%%%%%%%%
%%%%%%%%%%%%%%%%%%%%%%%%%%%%%%%%%%%%%%%%%%%%%
%%%%%%%%%%%%%%%%%%%%%%%%%%%%%%%%%%%%%%%%%%%%%

\section{Dataset description}\label{data_descr}

Our data source is Fermi~\cite{falciano2012}, an IBM Blue Gene/Q system run by CINECA, a consortium operating the largest data center in Italy. Fermi has 163,840 computing cores with a peak performance of 2.1 PFLOPS.
Its workload includes large-scale models and simulations for several academic projects, including 3D models of the cell network of the heart,  simulation of interaction between lasers and plasmas,  neuronal network simulations, models of nano-structures and complex materials.
Fermi is organized as 10 \emph{racks}, each with 2 \emph{mid-planes} of 16 \emph{node-boards} with 32 16-core nodes. Each mid-plane is powered by 18 \emph{bulk power modules} (BPM). Logging is based on standard Blue Gene/Q tools~\cite{lakner2013}. The \emph{Mid-plane Manager Control System} performs environmental monitoring, providing power and temperature logs. The \emph{Machine Controller} handles access to the hardware components and provides so-called \emph{Reliability, Availability and Serviceability} (RAS) logs.  Workload is extracted from the \emph{Portable Batch System} scheduler logs, using a custom tool by CINECA. Given that all data used in our analysis originate in logs from standard Blue Gene tools, we consider the information they contain to be correct.
 Table~\ref{table_datasets} summarizes the four datasets.

\begin{table}[!t]
\centering
\begin{tabular}{|p{1.9cm}|p{2.5cm}|p{2.3cm}|p{2.9cm}|p{1.9cm}|}
\hline
Dataset&Time~span (2014)&Time resolution&Component&Total  records\\
\hline
\hline
Power &28 Mar -- 25 Jul &5 min&Bulk Power Module&9,655,298\\
\hline
Temperature &23 Apr -- 25 Jul&15 min&Node-board&2,648,331\\
\hline
Workload&1 May -- 27 Jul&NA&System&78,128\\
\hline
RAS&23 Apr -- 25 Jul&NA&All&774,555\\
\hline
\end{tabular}
\vspace{1em}
\caption{Four datasets that are analyzed}
\label{table_datasets}
\vspace{-0.8cm}
\end{table}

\emph{Power} logs report input/output voltages and current levels for each BPM, with a 5-min resolution. By summing the input power levels over the different components, we obtained time series of power consumption for individual mid-planes, racks, and for the entire system. Power at the node-board scale cannot be reliably computed since 18 BPM power 16 node-boards (redundant system). \emph{Temperature} logs are reported by the node-board monitor (two sensors/node-board), with a 15-min resolution. From these we computed averaged time series at node-board, mid-plane, rack, and system scales. 

\emph{Workload} data consist of a list of jobs with date of completion, running time, number of cores, queue time, and queue class. Fermi uses six queues, with increasing job length and core count: \emph{serial} (on login nodes only), \emph{debug}, \emph{longdebug}, \emph{smallpar}, \emph{parallel} and \emph{bigpar}. Two other classes --- \emph{visual} and \emph{special} --- exist, with very few jobs reserved for dedicated users. We computed the CPU time per job and time series of total daily CPU time, number of cores, and queue time. The daily CPU time per queue class was also extracted. Since only the date of job completion (not the exact time) is available in the data, totals are approximate, yet they give a very good indication of the daily load at system scale. No load information at other scales (node-board, mid-plane, rack) was available.

\emph{RAS} logs consist of hardware and software events from all system components and are labeled \textsc{fatal}, \textsc{warn} or \textsc{info}, in decreasing order of severity. The dataset contains 163,134 \textsc{fatal}, 473,982 \textsc{warn}, and 137,438 \textsc{info} events.  For each event, the exact time and location are included. From these data, we computed the distribution of inter-event times at system scale and also time series of the number of events in each category at various time and space resolutions.

%%%%%%%%%%%%%%%%%%%%%%%%%%%%%%%%%%%%%%%%%%%%%
%%%%%%%%%%%%%%%%%%%%%%%%%%%%%%%%%%%%%%%%%%%%%
%%%%%%%%%%%%%%%%%%%%%%%%%%%%%%%%%%%%%%%%%%%%%
\section{Data analysis}\label{data_analysis}
Each dataset alone may provide useful insight into the functioning of Fermi, while an integrated analysis has even greater potential. Hence, in this section we first study each dataset individually, identifying and comparing their features, then we integrate them to study how metrics from different subsystems correlate. Pearson correlation coefficient is used across the paper to quantify correlations.

%%%%%%%%%%%%%%%%%%%%%%%%%%%%%%%%%%%%%%%%%%%%%
%%%%%%%%%%%%%%%%%%%%%%%%%%%%%%%%%%%%%%%%%%%%%
%%%%%%%%%%%%%%%%%%%%%%%%%%%%%%%%%%%%%%%%%%%%%

\subsection{Individual datasets}

%%%%%%%%%%%%%%%%%%%%%%%%%%%%%%%%%%%%%%%%

%%%REVIEWER 4: Further, the various data plots are hard to compare since they are on different pages. A rearrangement such that they are next to each other (especially for power and temperature) would be useful.
%%AS : I put these two plots on the same page.

%%%%%%%%%%%%%%%%%%%%%%%%%%%%%%%%%%%%%%%%

\begin{figure}[!t]
\centering
\vspace{-0.3cm}
\subfloat[System power]{\includegraphics[width=2in]{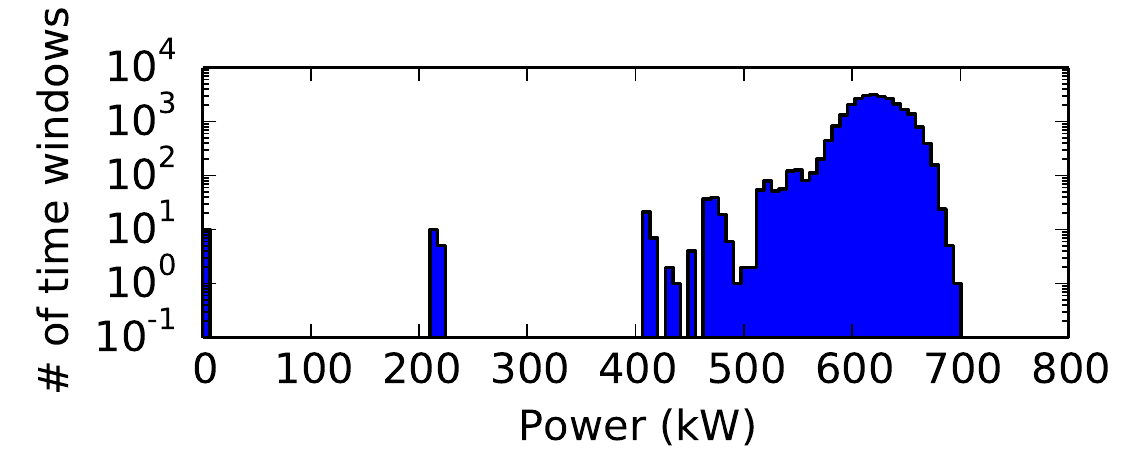}%
\label{fig_power_hist_system}}
\subfloat[Mid-plane power]{\includegraphics[width=2in]{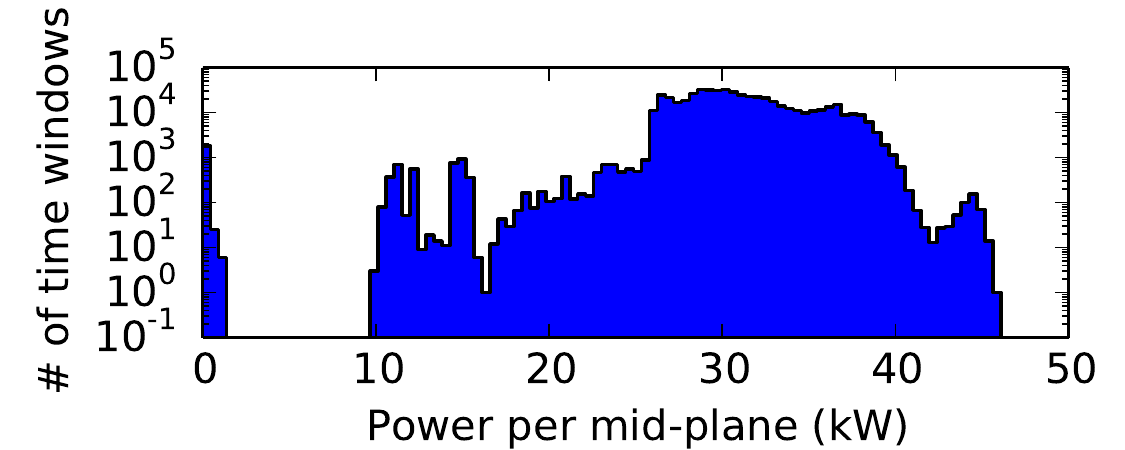}%
\label{fig_power_hist_midplane}}
\\
\vspace{-0.4cm}
\subfloat[Rack correlations]{\includegraphics[width=2in]{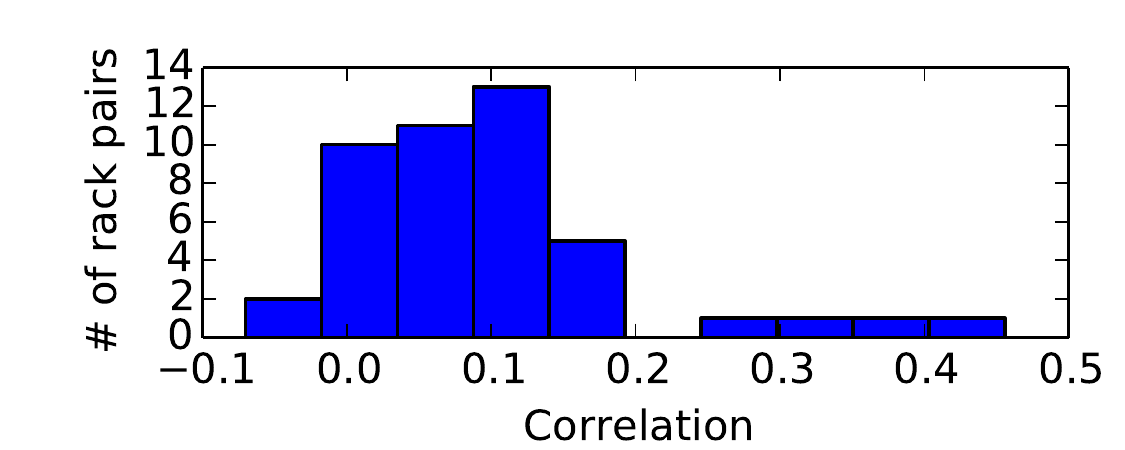}%
\label{fig_power_corr_rack}}
\subfloat[Mid-plane correlations]{\includegraphics[width=2in]{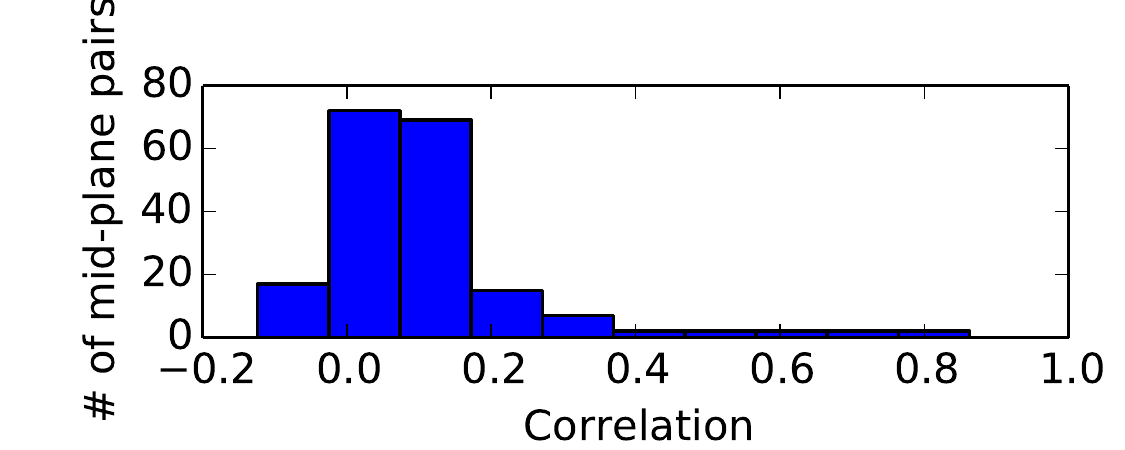}%
\label{fig_power_corr_midplane}}
\caption{Distribution of total power, sampled every 5 min, at system and mid-plane scale and of power correlations between racks and mid-planes.}
\label{fig_power}
\vspace{-0.5cm}
\end{figure}

\begin{figure}[!b]
\centering
\vspace{-0.8cm}
\subfloat[System]{\includegraphics[width=2in]{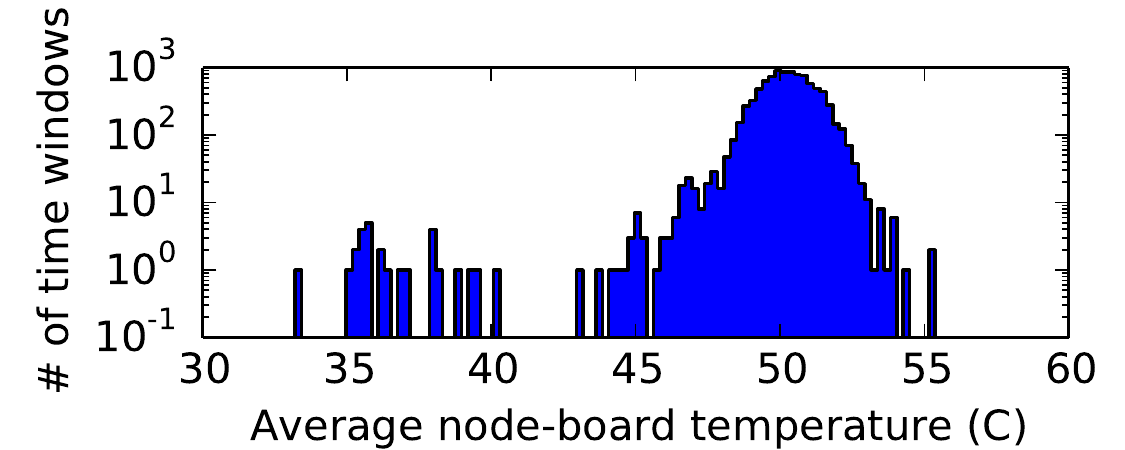}%
\label{fig_temp_hist_system}}
\subfloat[Node-board]{\includegraphics[width=2in]{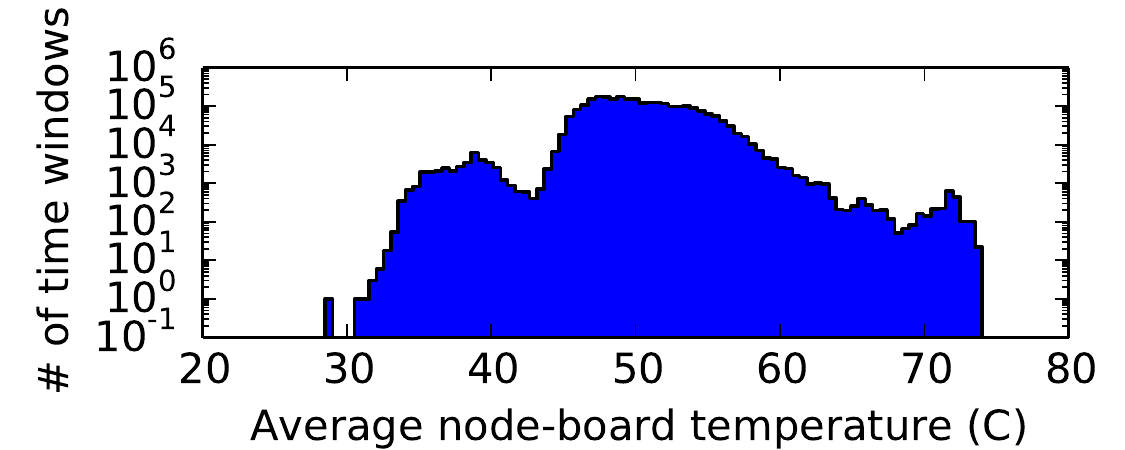}%
\label{fig_temp_hist_nodeboard}}
\\
\vspace{-0.4cm}
\subfloat[Rack]{\includegraphics[width=2in]{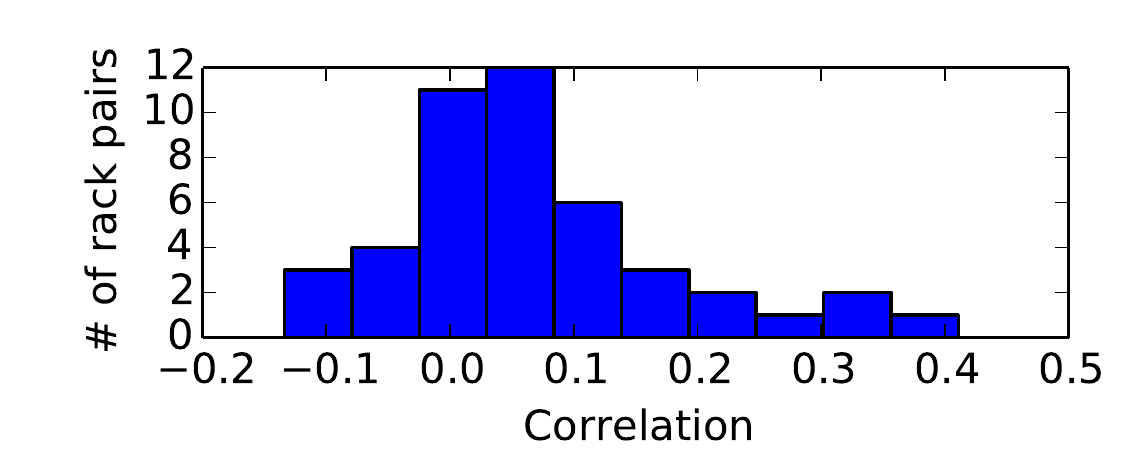}%
\label{fig_temp_corr_system}}
\subfloat[Node-board]{\includegraphics[width=2in]{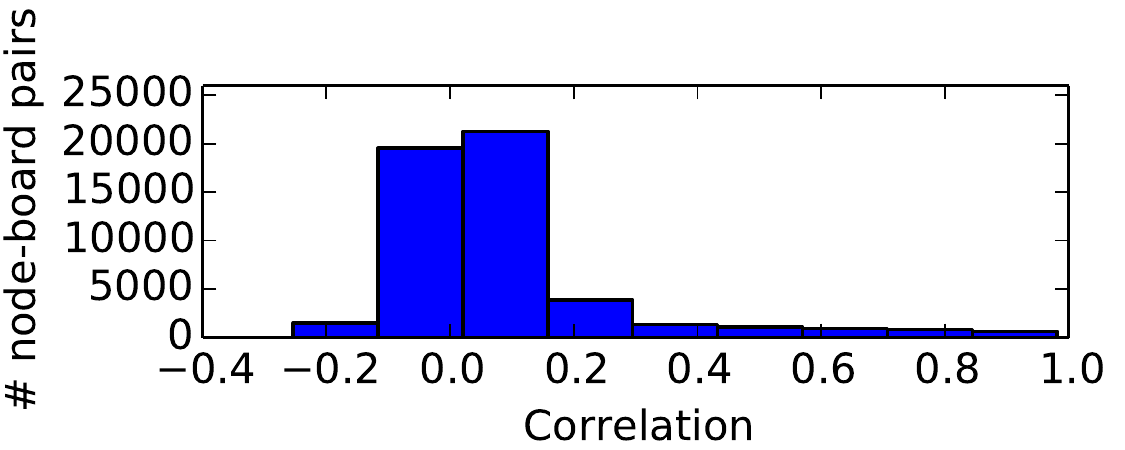}%
\label{fig_temp_corr_nodeboard}}
\caption{Distribution of average temperatures, at system and node-board scale, and of temperature correlations among racks and node-boards.}
\label{fig_temp}
\vspace{-0.3cm}
\end{figure}

{\bf Power logs.} The specifications for a Blue Gene/Q system declare the typical power consumption to be around 65kW, with a maximum of 100kW per rack~\cite{milano2013}.
However, real consumption varies depending on system load and state of components (e.g., how many nodes are up). Fig.~\ref{fig_power}a-b displays the distribution of power consumption sampled at 5-min intervals, at system and mid-plane scale. Distributions are centered around the official average values: 650kW at the system scale (10 racks) and 32.5kW at the mid-plane scale, confirming the specifications.  Moving from higher to lower scale, the distribution becomes broader. While total consumption is mostly between 50kW and 70kW per rack, with a bell-shaped distribution, for individual mid-planes additional peaks emerge with some showing power consumptions up to 46kW, but also frequent values under 20kW. Similar results were obtained at rack scale. This shows that power consumption is very heterogeneous, which needs to be taken into account for modeling. Indications are that while predicting overall system power might be easier due to greater stability in time, finer grained predictions at mid-plane scale might produce more accurate results.

%%%%%%%%%%%%%%%%%%%%%%%%%%%%%%%%%%%%%%%%

%%REVIEWER 2:
%%%Analysis: A number of unsubstantiated claims are made in the paper. For example, the authors do not find a strong correlation between components with respect to power and temperature, as might be expected.  They claim, without proof, the reason for this is poor load balancing in the workloads run.  Given the limitations in there data collection infrastructure, they need to provide some more complete support for this claim.

%%REVIEWER 1:
%%%Really would like to see a bit more discussion about the lack of correlation between levels of the power train.  The comment about poor design of applications leaves me wanting more details.

%%AS:
%%%We do not have details about applications and where they run, so we cannot give more details for these two comments. I added a statement to mention this and added it to future work in the last section.

It is interesting to see if power correlates across different components (racks or mid-planes). Traditional load balancing algorithms try to even out the work performed by different processing elements, and power increases with load, so we would expect power to be correlated across different system components under heavy load. Fig.~\ref{fig_power}c-d shows  correlations of power consumption between rack pairs, and between mid-plane pairs. At both scales, correlations are in general very low. Only a few mid-plane pairs have correlation values above 0.5. As we will see later, the observed system load is generally high. The lack of strong correlation for power consumption among components could be interpreted as an effect of energy-aware scheduling~\cite{valentini2013}, yet, this is not the case here since Fermi uses the native IBM LoadLeveler scheduler which is not optimized for power. A different explanation for the weak correlations could be poor design of the applications running on the system: if synchronization requires some program threads to wait, these will keep the nodes occupied but without using them fully. \revised{However, given the coarse resolution of the workload dataset, this hypothesis cannot be tested with the current data.}

%%%%%%%%%%%%%%%%%%%%%%%%%%%%%%%%%%%%%%%%%%%%%
%%%%%%%%%%%%%%%%%%%%%%%%%%%%%%%%%%%%%%%%%%%%%
%%%%%%%%%%%%%%%%%%%%%%%%%%%%%%%%%%%%%%%%%%%%%

{\bf Temperature logs.} 
Fig.~\ref{fig_temp}a-b shows histograms of average temperatures sampled every 15 min. For the overall system, with few exceptions, the distribution is again bell-shaped and narrow with one mode around 50$^{\circ}$C. As we zoom in at node-board scale (the lowest available in the data), the distribution becomes again wider with additional peaks appearing at very high and very low temperatures. Individual node-boards can reach up to 75$^{\circ}$C, significantly greater than the system average. Similar results were obtained at intermediate scales (rack and mid-plane). This again shows how the system appears to behave differently at different scales, with greater heterogeneity in time at the finer-grained logs. For temperature correlations among different components of the same type (Fig.~\ref{fig_temp}c-d), a pattern similar to power consumption is observed. With very few exceptions, temperature exhibits low correlation across components. Results are consistent across all scales (including mid-plane not shown here). In terms of thermal isolation, this is good news, since having one hot node-board does not imply surrounding node-boards are hot as well. Yet, the fact that power consumption showed a similar pattern, this can be additional evidence that workload is not well balanced or applications need improvement.

%%%%%%%%%%%%%%%%%%%%%%%%%%%%%%%%%%%%%%%%%%%%%
%%%%%%%%%%%%%%%%%%%%%%%%%%%%%%%%%%%%%%%%%%%%%
%%%%%%%%%%%%%%%%%%%%%%%%%%%%%%%%%%%%%%%%%%%%%

\begin{figure}[!t]
\centering
\vspace{-0.2cm}
\subfloat[CPU time/job - multiples of $10^9$s]{\includegraphics[width=2in]{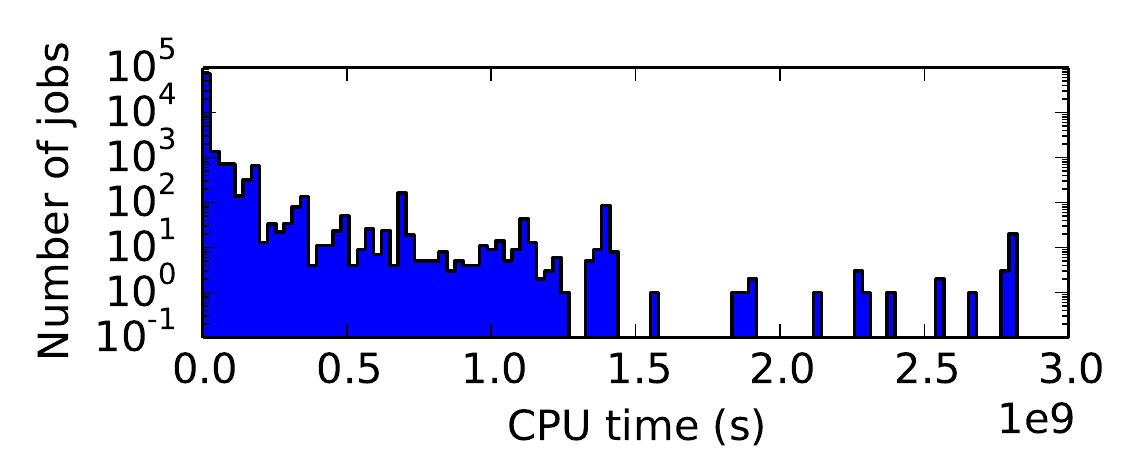}%
\label{fig_job_cput}}
\subfloat[Running time/job]{\includegraphics[width=2in]{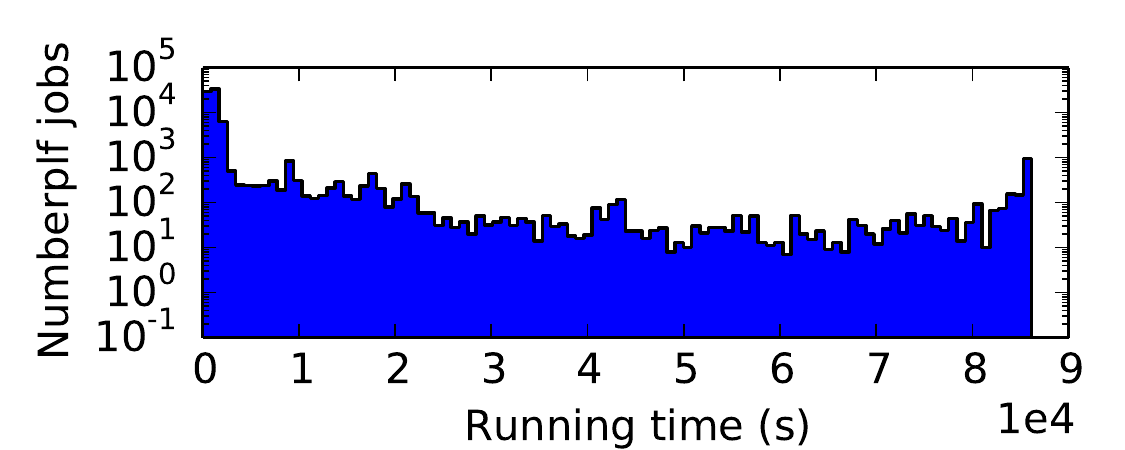}%
\label{fig_job_qt}}
\\
\vspace{-0.4cm}
\subfloat[Core count/job]{\includegraphics[width=2in]{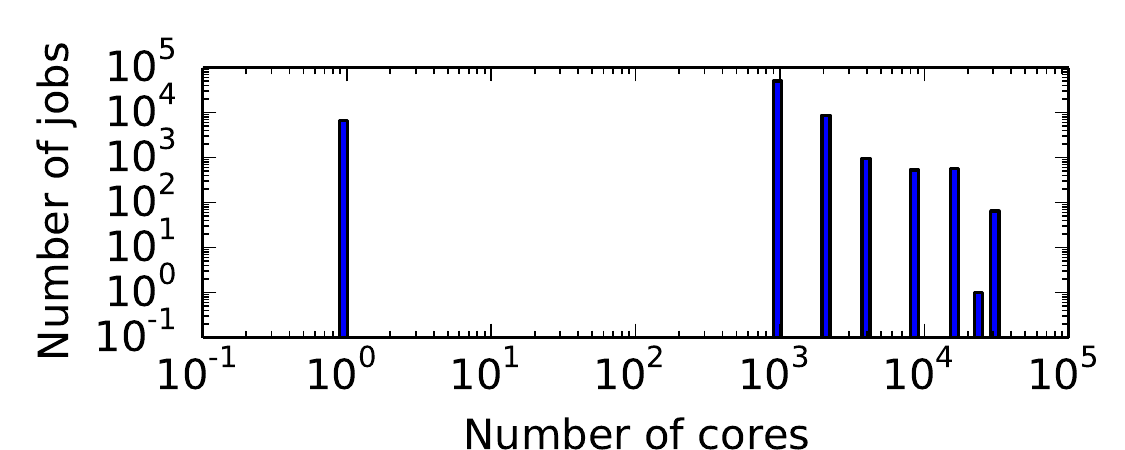}%
\label{fig_job_cores}}
\subfloat[Daily CPU time (normalized) ]{\includegraphics[width=2in]{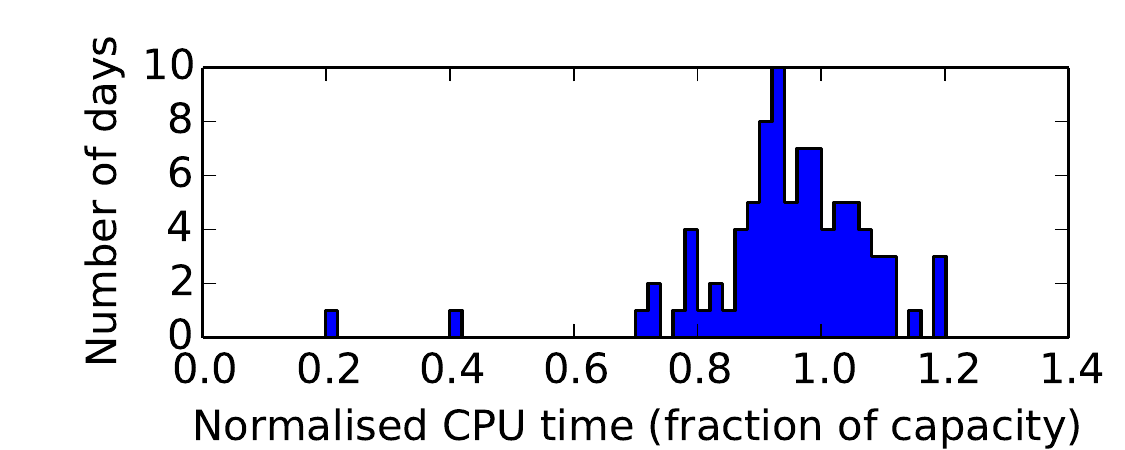}%
\label{fig_load}}
\caption{Workload structure: distribution of CPU time, running time, number of cores per job, and CPU time consumed by jobs completed on the same day (normalized by the overall capacity of Fermi, which is 14,069,376,000 s/day).}
\label{fig_job}
\vspace{-0.6cm}
\end{figure}

{\bf Workload logs.}
An important question in terms of workload regards the types of jobs submitted to the system. Fig.~\ref{fig_job}a-c displays the distribution of several job attributes: CPU time, running time, and number of cores used. In terms of time requirements, jobs are very heterogeneous as evidenced by a  long-tailed CPU time distribution, with a few very heavy jobs and many short jobs present.  Effective running times are bimodal, with many short jobs and many long jobs (all running times under 24 hours), and slightly fewer medium-length jobs. The number of cores per job is less heterogeneous, with only eight different values present, most jobs using over 100 cores and up to 32,768. So, in general, jobs are highly parallel. Out of all 78128 jobs submitted, only  $\sim$75\% were started (running times $>0$) and only those will be used in the subsequent sections.

The structure of the workload data enables analysis of patterns in time only at system scale and 24-hour resolution. Fig.~\ref{fig_job}d shows total CPU time for all jobs \emph{completed} each day, normalized by the overall system capacity. This does not represent the exact system load for that day, but it still is a very good indication. The data contain only the date of job completions not the exact time, making it impossible to compute how many hours each job ran in a given day --- jobs completed on one day could have been started the previous day. This is why some days reach capacity exceeding 100\%. Again, roughly a bell-shaped distribution is observed, with a mean around 94\% usage, indicating very high load levels.

\begin{figure}[!t]
\centering
\vspace{-0.2cm}
\subfloat[Inter-event times]{\includegraphics[width=2in]{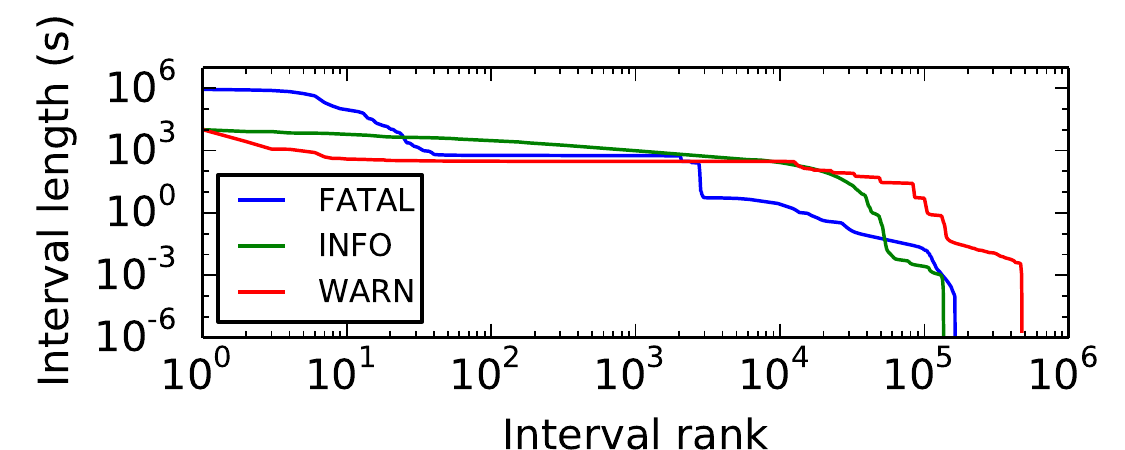}%
\label{fig_RAS_interevent}}
\subfloat[Daily events]{\includegraphics[width=2.1in]{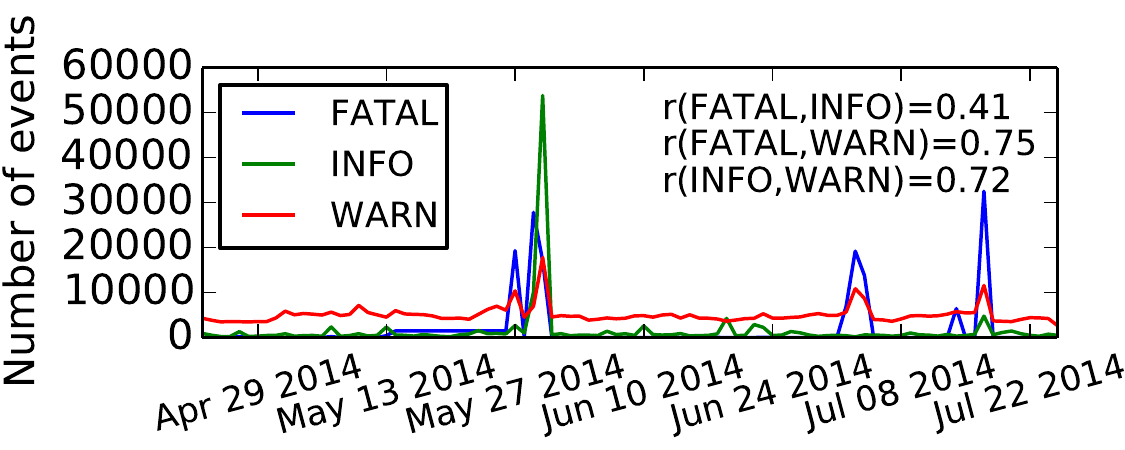}%
\label{fig_RAS_daily}}
\\
\vspace{-0.4cm}
\subfloat[Rack correlation]{\includegraphics[width=2in]{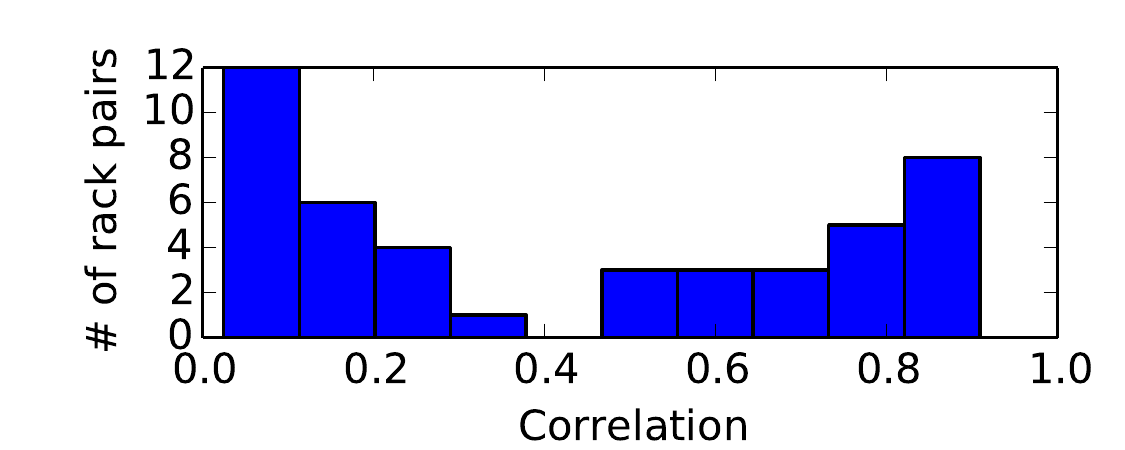}%
}
\subfloat[Node-board correlation]{\includegraphics[width=2in]{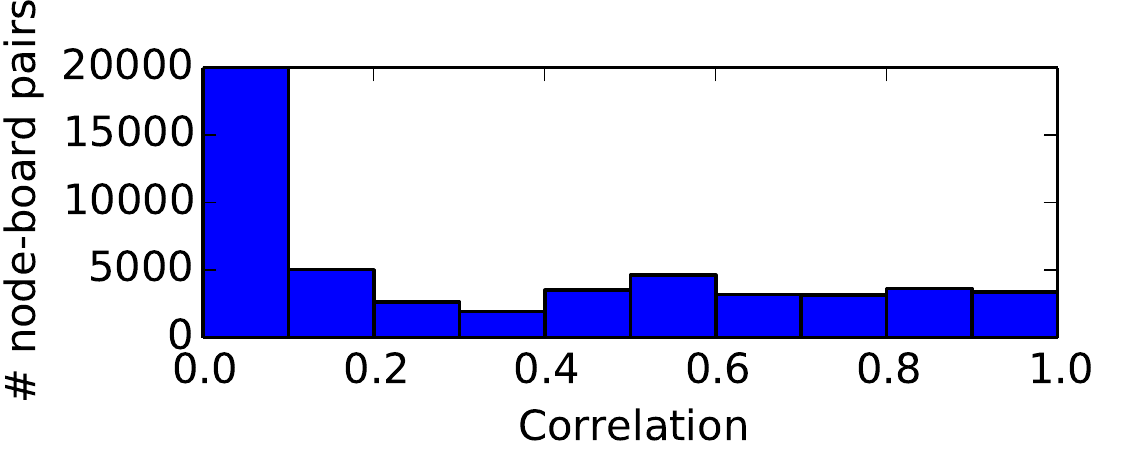}%
}
\caption{(a) Inter-event times for the three event categories. Intervals between events were ranked in descending order. For each interval, the $x$ axis shows its rank and the $y$ axis its value. (b) Total daily number of RAS events. Correlation  of \textsc{fatal} events for rack pairs (c) and node-board pairs (d).}
\label{fig_RAS}
\vspace{-0.6cm}
\end{figure}

%%%%%%%%%%%%%%%%%%%%%%%%%%%%%%%%%%%%%%%%%%%%%
%%%%%%%%%%%%%%%%%%%%%%%%%%%%%%%%%%%%%%%%%%%%%
%%%%%%%%%%%%%%%%%%%%%%%%%%%%%%%%%%%%%%%%%%%%%
{\bf RAS logs.}
The inter-event times at system scale, for the three event types, do not appear to follow a known distribution (Figure~\ref{fig_RAS_interevent}). \textsc{fatal} events show a few very large and many very small intervals, indicating a pattern with spikes of events in short periods of time with large breaks between them. \textsc{info} and \textsc{warn} events are more evenly spread in time, missing the very large inter-event times, and having a smaller fraction of very short intervals.

Fig.~\ref{fig_RAS_daily} displays the time-series of daily number of events in each category and their relative correlations. \textsc{warn} and \textsc{info} events are more common daily, whereas \textsc{fatal} events come in spikes and appear in only a few of the monitored days.  
The 4 larger spikes in \textsc{fatal} events correspond to issues related to the BPMs which caused shutdown of the entire system several times between 27/05 and 30/05 and shutdown of rack R30 on 04/07 and 17/07. 
Daily \textsc{info} and \textsc{warn} events are highly correlated, and so are \textsc{warn} and \textsc{fatal} events. However \textsc{info} and \textsc{fatal} events seem to appear together less frequently. This could mean that \textsc{info} events could be useful to predict \textsc{warn} events while \textsc{warn} events could predict \textsc{fatal} events at this time resolution. Hence, considering both \textsc{info} and \textsc{warn} events to predict \textsc{fatal} events could facilitate longer prediction lead time.  

A different question is whether events correlate across different components. Fig.~\ref{fig_RAS}c-d shows the distribution of correlations between rack and node-board pairs for \textsc{fatal} events. Similar results were obtained for the other events and at mid-plane scale. Unlike power and temperature, \textsc{fatal} events have higher correlation across components, with a significant number of pairwise correlations larger than 0.5. This indicates that failures may propagate across components. We studied for various \textsc{fatal} event types the number of different components (node-boards, power modules, etc.) affected in 5-min windows. We found that most event occurrences do involve a large number of components, sometimes up to a few hundred. So, when trying to predict component failures, one needs to take into account not only their individual behavior, but also that of the others. The way failures propagate can also give indications of the possible causes (e.g., a faulty job running on all components) and enable their automatic identification.

%%%%%%%%%%%%%%%%%%%%%%%%%%%%%%%%%%%%%%%%%%%%%
%%%%%%%%%%%%%%%%%%%%%%%%%%%%%%%%%%%%%%%%%%%%%
%%%%%%%%%%%%%%%%%%%%%%%%%%%%%%%%%%%%%%%%%%%%%
\subsection{The big picture}\label{sec_corr}

Individual datasets have shed some light into the functioning of the Fermi system, and correlations between components. Here we integrate the four datasets to uncover further correlations between the different components and logs.

\begin{figure}[!t]
\centering
\vspace{-0.2cm}
\includegraphics[width=3.9in]{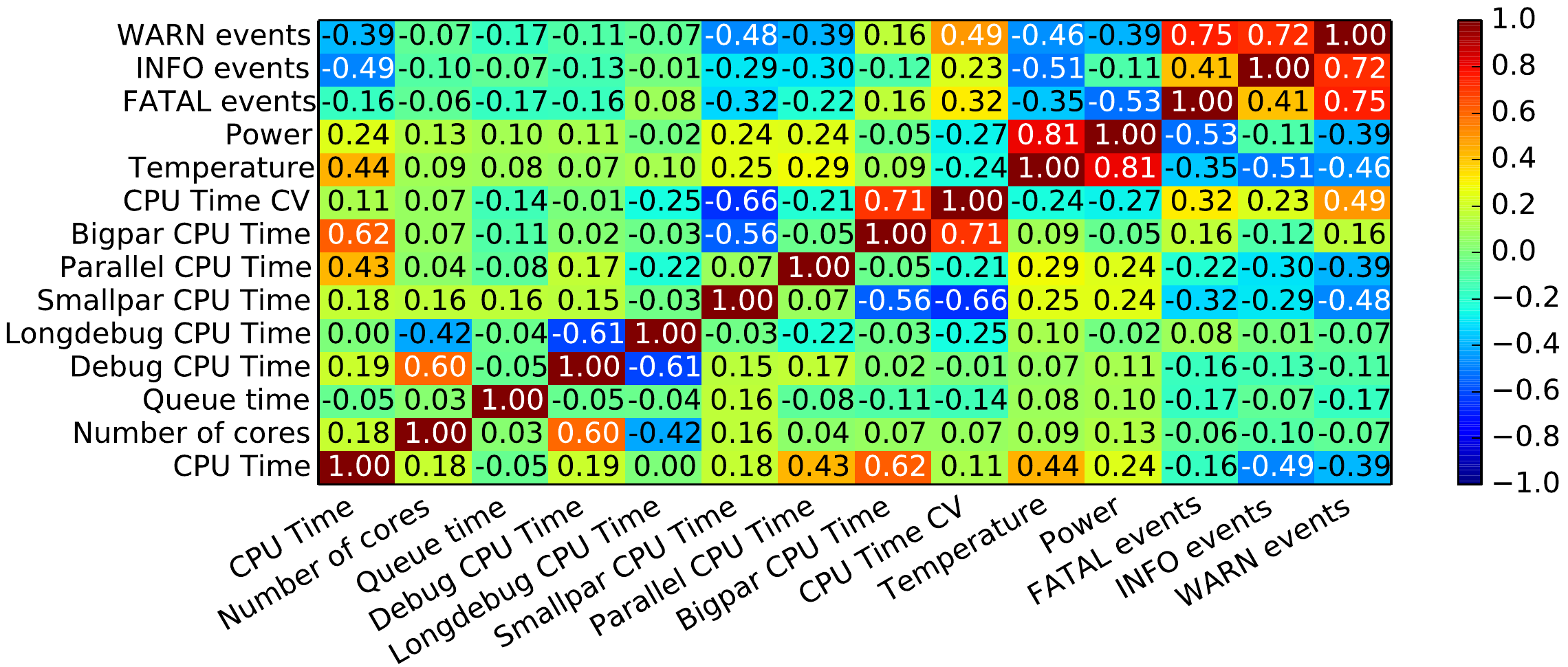}
\caption{Correlation between datasets at 24-hour resolution for the overall system.}
\label{fig_all_corr_day}
\vspace{-0.2cm}
\end{figure}

A first analysis looks at different measures for the overall system for 24-hour time windows. Figure~\ref{fig_all_corr_day} shows all pairwise correlations between several time series datasets. We note strong correlation between temperature and power, confirming what has been observed in other systems as well~\cite{Bartolini2014}. In terms of workload, total daily CPU time, number of cores, and queue time are included. These do not appear to correlate among themselves, while CPU time is the only one among the three that does correlate with other datasets, although only moderately.  Specifically, positive correlation with the temperature is present, so the system does show thermal symptoms of working harder under a high workload. A negative correlation with RAS events also exists, which is somewhat counterintuitive: one would expect more events to appear when the system works harder. However, it is quite possible for large numbers of RAS events to have resulted in system failure, which in turn resulted in fewer completed jobs for that day, explaining the negative correlation. In fact, a closer analysis of the data shows that, in general, a system shutdown (signaled by long periods of missing data in the trace) is preceded by fatal events. In some situations, events may appear also at system restore, which could be due to operator interventions made while the system was down. A negative correlation also appears between power/temperature and RAS events, again rather counterintuitively and due to the same factors as before. So, when trying to predict power consumption or \textsc{fatal} events, one needs to take into consideration the negative dependence.

\begin{figure}[!b]
\vspace{-0.5cm}
\centering
\includegraphics[width=3in]{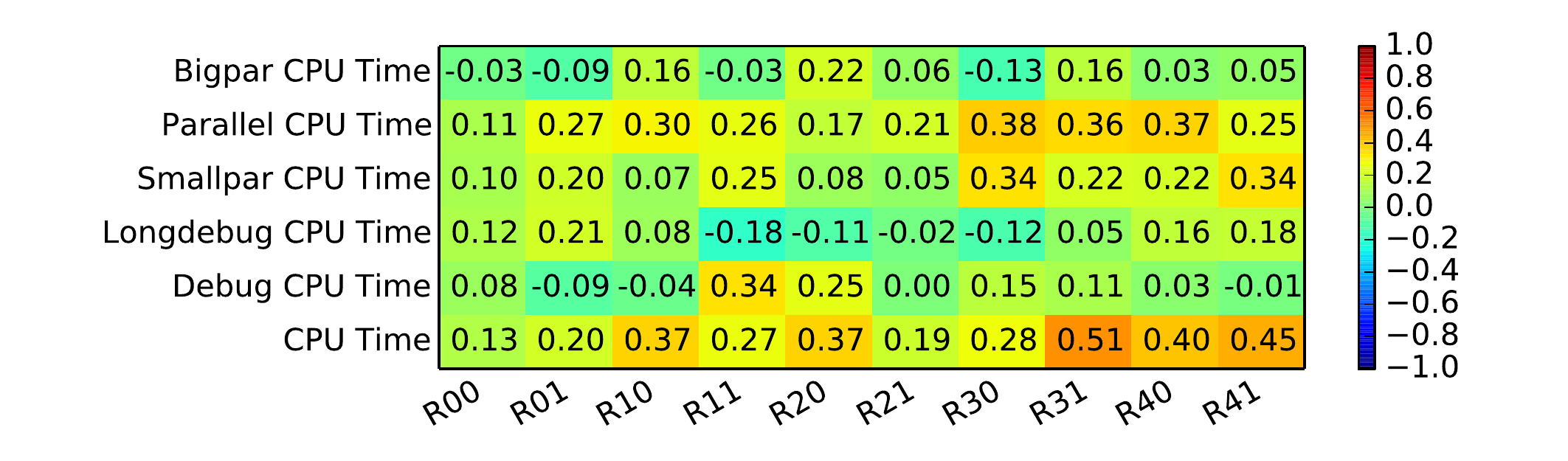}
\caption{Correlation between CPU time and power for the 10 racks (R00-R41).}
\label{fig_rack}
\end{figure}

The data do not show any correlation between overall workload and power, however correlation could depend on the job class (queue). So we analyzed (same Fig.~\ref{fig_all_corr_day}) the daily CPU Time per job class and also the coefficient of variation (CV) of the total CPU Time across the classes. A higher CV means a more unbalanced workload across the queues. The negative correlation between CPU time and RAS events is present for individual queues as well, with strongest effect for \emph{smallpar} jobs. CPU Time CV displays some positive correlation to \textsc{warn} events, which means that heterogeneity in terms of jobs per queue can be a factor leading to \textsc{warn} events. However, even at this scale, no link between workload and power consumption can be found (we also explored other measures, such as job count, core count, queue time per class, with similar results). This suggests that the way workload is distributed on components is important to understand power in this system. Higher correlations might be obtained by zooming in at rack, mid-plane or node-board scales. We can do this for power, but not for workload due to the structure of our dataset. Fig.~\ref{fig_rack} shows how CPU Time correlates with power consumption per rack. Indeed, higher correlations do appear, indicating structure is  important, but still more detailed workload data is required. This suggests the need for changes in the structure of workload logs for Fermi and improving system logging practice, in order to see exactly at which scale correlations appear.

\begin{figure}[!t]
\centering
\vspace{-0.3cm}
\subfloat[System]{\includegraphics[width=1.5in]{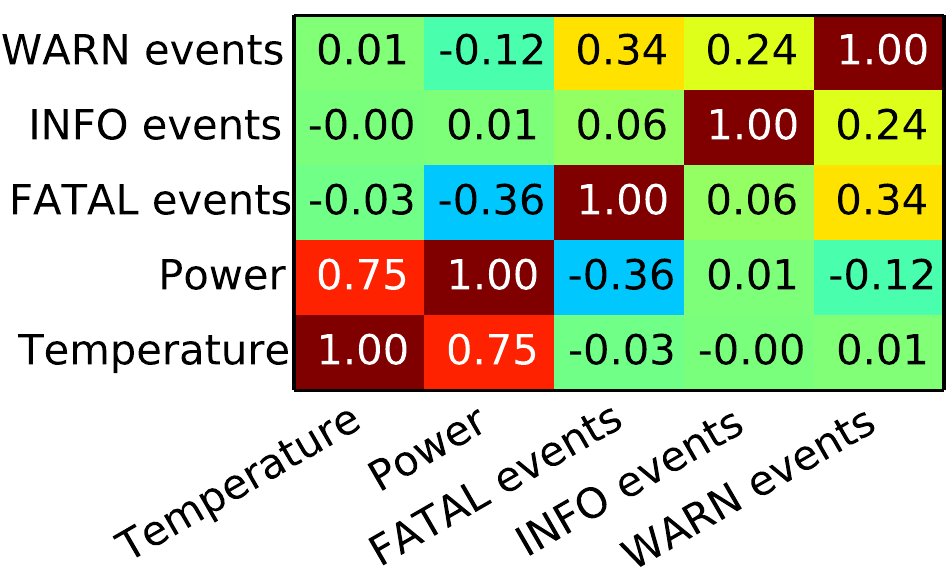}%
\label{fig_all_corr_min}}
\subfloat[Rack]{\includegraphics[width=1.5in]{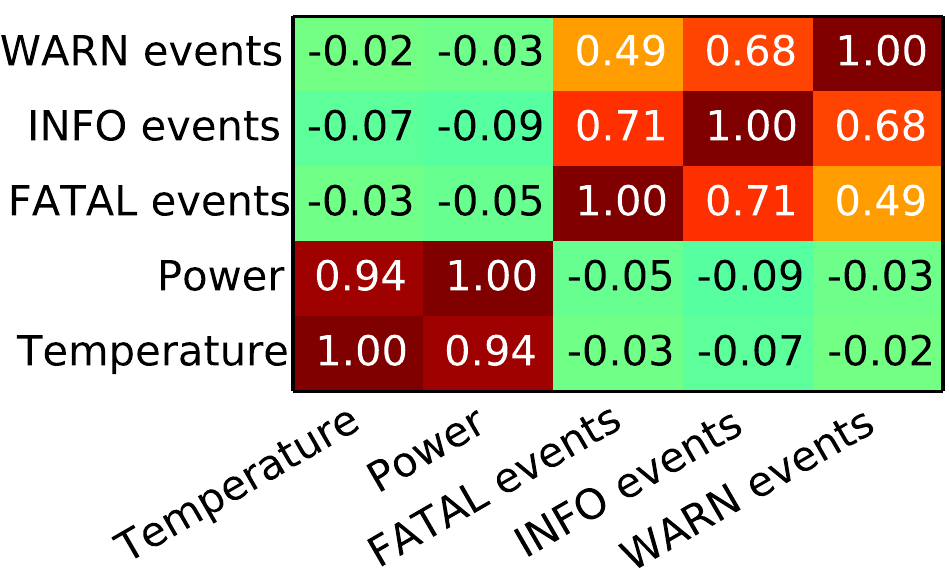}%
\label{fig_corr_r}}
\subfloat[Mid-plane]{\includegraphics[width=1.8in]{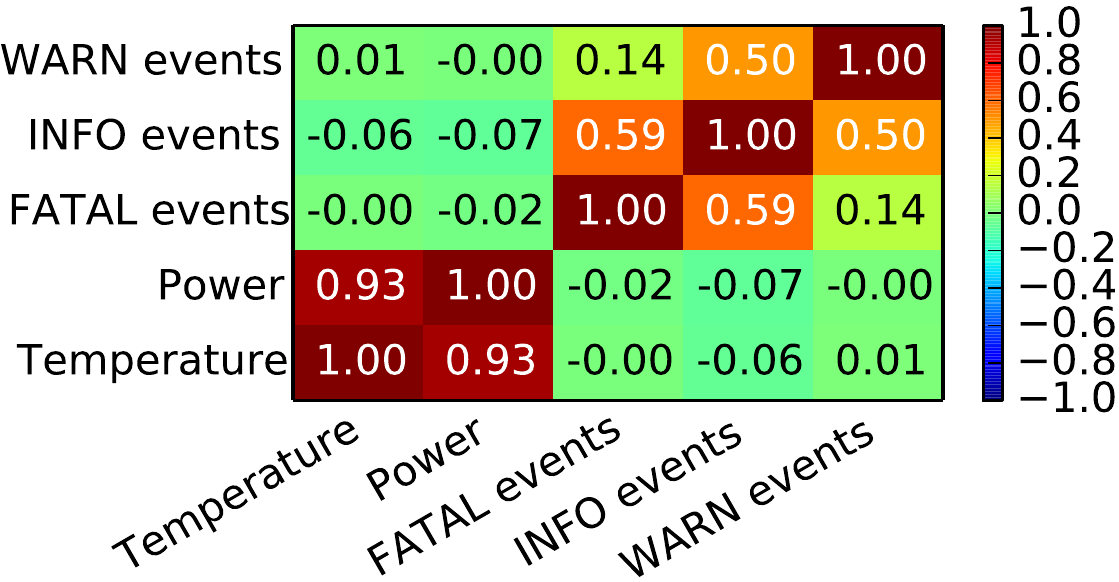}%
\label{fig_corr_mp}}
\caption{Correlation between datasets at 3 different system scales.}
\label{fig_corr}
\vspace{-0.5cm}
\end{figure}

In a second analysis, the resolution of the data was increased to 5 min. Correlations at system, rack, and mid-plane scales are shown in Fig.~\ref{fig_corr}. Due to its coarse time structure, the workload data was excluded. Power and temperature correlation grows with increased time and space resolution. This suggests that for predictions, using only one of the two features might suffice, which is good news since power logs at node-board scale are not available. 
However, to account for the possibility that temperatures are affected by cooling issues, power should still be monitored, even if not to be used as a modeling feature, but to check that assumed correlation is correct. A sudden decrease in correlation could also flag cooling anomalies.
The negative correlation between temperature/power and RAS events is  maintained, albeit at a lower value, only at system scale between power and \textsc{fatal} events.This can be again explained by the existence of periods of system shutdown before or after events. So, correlations will be high over 24 hours, but for 5-min windows, only \textsc{fatal} events are correlated with power (temperatures take longer to drop, while other event types could have appeared much earlier). Between RAS events, correlations are higher within racks, indicating that propagation of errors might be strongest at the rack scale.

%%%%%%%%%%%%%%%%%%%%%%%%%%%%%%%%%%%%%%%%%%%%%
%%%%%%%%%%%%%%%%%%%%%%%%%%%%%%%%%%%%%%%%%%%%%
%%%%%%%%%%%%%%%%%%%%%%%%%%%%%%%%%%%%%%%%%%%%%
\section{Related work}\label{related}

Log analysis for characterization of large computing infrastructures has been the focus of numerous recent studies.  The release of two Google workload traces has triggered a flurry of analysis activity. General statistics, descriptive analyses, and characterization studies~\cite{Liu2012,Reiss2012} have revealed higher levels of heterogeneity when compared to grid systems~\cite{Di2012}. Some modeling work has also appeared based on these data~\cite{Wang2011,Balliu2014,sirbu2015}. While they have provided important insight into Google clusters, focusing only on workload aspects of the system has been limiting.  To be effective, it is essential to integrate data from different components and sources. Other traces have also been studied in the past~\cite{Chen2012}, and tools for their analysis developed~\cite{Javadi2013}, but again concentrating on a single data type.  Here we perform similar analyses for a Blue Gene/Q system but from several viewpoints: workload, RAS, power, and temperature\revised{, providing a more complete picture of the system under study.}

RAS logs from IBM Blue Gene systems have been included in several earlier studies. In~\cite{Dudko2012} prediction of \textsc{fatal} events in a Blue Gene/Q machine is attempted while an earlier study of a Blue Gene/L installation is~\cite{Liang2007}. Both compare several classification tools (SVM, customized KNN, ANN, feature selection, rule-based models). These predictive studies look only at RAS events, while adding further data from other system components could improve prediction accuracy significantly, as noted by the authors themselves. In this paper we provide the first step towards such an analysis, where we perform descriptive analytics mandatory before any prediction can be attempted.

Some integration is performed in a very recent study from Google that models Power Usage Effectiveness using thermal information (temperatures, humidity, etc) and overall system load, using an Artificial Neural Network~\cite{Gao2014}. Another recent development in this direction is a novel monitoring system~\cite{Bartolini2014} designed for a hybrid HPC platform. In this study, several types of data including workload, power, chiller, and machine status are recorded. In principle, these data could be used for future predictive and modeling studies, \revised{but they have not been initiated in the reported study.}
The OVIS project has also developed an integrated monitoring platform called the ``Lightweight Distributed Metric Service'' \cite{Agelastos2014}, recording various system metrics for optimization of application performance. The platform has been tested on several systems, \revised{but again steps towards a descriptive and predictive analytics for these data are still missing.}

%%%%%%%%%%%%%%%%%%%%%%%%%%%%%%%%%%%%%%%%%%%%%
%%%%%%%%%%%%%%%%%%%%%%%%%%%%%%%%%%%%%%%%%%%%%
%%%%%%%%%%%%%%%%%%%%%%%%%%%%%%%%%%%%%%%%%%%%%
\section{Discussion and conclusions}\label{discussion}

Given the need for a holistic analysis of large computing infrastructures, this paper has presented a characterization study conducted with four datasets describing different subsystems of an IBM Blue Gene/Q installation. Temperature, power consumption, workload, and RAS logs were studied independently to characterize the system and then together to identify correlations between datasets. 

The results obtained from correlation analysis will serve as a guideline for a future study aiming to predict in advance \textsc{fatal} RAS events based on the rest of the data. One possibility would be predicting, for each node-board, the number of \textsc{fatal} events in the next 24 hours. 
Alternatively, based on the number of events, we can define discrete failure classes (e.g., \textsc{none}, \textsc{few}, \textsc{many}) to be predicted. We have compiled a set of possible features that may be suitable for this predictive task (Table \ref{table_features}). These cover all datasets with various time resolutions at node-board and system scale. 

The first two features are suggested by the fact that power and temperatures are highly correlated. Temperatures can be used a a proxy for power, so that the higher space resolution is employed and the number of features is decreased. This only as long as correlation between temperature and power is high. A decrease in correlation will signal an anomaly, even if the proxy is no longer valid. Large temperature correlations across node-boards could also signify anomalies, since node-board temperatures were uncorrelated in our data. Workload related features are limited to daily CPU time per queue and coefficient of variation across queues, which showed highest correlation with other datasets. Features monitoring all types of RAS events at node-board level account for correlation across RAS event types, while those at system level are justified by correlations across node-boards and propagation of errors. Since prediction is aimed for 24-hour periods, we use event values computed over the same time, but also deviations, to account for varying inter-event patterns. Finally, correlations between power (or temperature at node-board scale) and events should be monitored since large negative correlation could signal component failure.
Even if indications are that the features listed will prove important for prediction, final evaluation of the feature set will be performed during the future predictive study itself.

\begin{table}[!t]
\centering
\vspace{-0.6cm}
\begin{tabular}{|p{8cm}|p{1.5cm}|p{1.8cm}|}
\hline
\bf{Feature}& \bf{Period}&\bf{Scale}\\
\hline
Temperature average and standard deviation & 6h &node-board \\
\hline
Correlation between temperature and power & 6h  &mid-plane \\
\hline
Temperature correlation between node-boards &6h&node-board\\
\hline
CPU time per queue & 24h &system\\
\hline
CPU time coefficient of variation across queues & 24h &system\\
\hline
Number of \textsc{warn}, \textsc{info}, and \textsc{fatal} events & 24h &node-board\\
\hline
Stdev of number of \textsc{warn}, \textsc{info}, and \textsc{fatal} events & 24h &node-board \\
\hline
Number of \textsc{warn}, \textsc{info}, and \textsc{fatal} events & 24h &system\\
\hline
Stdev of number of \textsc{warn}, \textsc{info}, and \textsc{fatal} & 24h &system \\
\hline
Correlation between temperature and event count &6h&node-board\\
\hline
Correlation between power and event count &6h&system\\
\hline
\end{tabular}
\vspace{1em}
\caption{Possible feature set for prediction of \textsc{fatal} RAS events.}
\label{table_features}
\end{table}

Besides identifying important features, our analysis has also indicated directions for improvement in terms of data collection. Workload data in particular proved to be insufficient for our goals, so we could identify few relations to the other datasets. In the future, at least timestamps for job completion as well as job placement should be included. \revised{This additional information will enable analyzing the causes of lack of power correlation across components.} Power monitoring was coarse in terms of space resolution, however more data could be extracted from the node-board power rails. Temperatures, on the other hand, could be logged at 5-min intervals rather than 15. We are aiming at prediction with long lead time, so the 15-min interval may be sufficient for applying the model, however finer granularity would allow for more refined training data. In the future we will also use data external to the computing infrastructure, such as the water and air cooling systems, together with data outside the data center, e.g. weather and seismic activity. Cross-correlations will also be investigated, resulting in further features to be added to the proposed set.

%%%%%%%%%%%%%%%%%%%%%%%%%%%%%%%%%%%%%%%%%%%%%
%%%%%%%%%%%%%%%%%%%%%%%%%%%%%%%%%%%%%%%%%%%%%
%%%%%%%%%%%%%%%%%%%%%%%%%%%%%%%%%%%%%%%%%%%%%
\section{Acknowledgments}
We are grateful to the HPC team at CINECA for sharing with us the log data related to the Fermi system and for helpful discussions.  

%%%%%%%%%%%%%%%%%%%%%%%%%%%%%%%%%%%%%%%%%%%%%
%%%%%%%%%%%%%%%%%%%%%%%%%%%%%%%%%%%%%%%%%%%%%
%%%%%%%%%%%%%%%%%%%%%%%%%%%%%%%%%%%%%%%%%%%%%
\bibliographystyle{splncs03}
\bibliography{refs}

\end{document}